# Automated Multi-Source Debugging and Natural Language Error Explanation for Dashboard Applications

Devendra Tata
Principal Software Engineer
Palo Alto Networks
dtata@paloaltonetworks.com

Mona Rajhans
Senior Manager, Software Engineering
Palo Alto Networks
mrajhans@paloaltonetworks.com

**Abstract -** Modern web dashboards and enterprise applications increasingly rely on complex, distributed microservices architectures. While these architectures offer scalability, they introduce significant challenges in debugging and observability. When failures occur, they often manifest as opaque error messages to the end-user (e.g., "Something went wrong"), masking the underlying root cause which may reside in browser-side exceptions, API contract violations, or server-side logic failures. Existing monitoring tools capture these events in isolation but fail to correlate them effectively or provide intelligible explanations to non-technical users. This paper proposes a novel system for Automated Multi-Source Debugging and Natural Language Error Explanation. The proposed framework automatically collects and correlates error data from disparate sources (browser, API, server logs), validates API contracts in real-time, and utilizes Large Language Models (LLMs) to generate natural language explanations. This approach significantly reduces Mean Time to Resolution (MTTR) for support engineers and improves the user experience by transforming cryptic error codes into actionable insights.

## I. Introduction

The evolution of web applications from monolithic structures to distributed microservices has revolutionized software development, enabling rapid deployment and independent scaling. However, this shift has also fragmented the debugging landscape. A single user action on a dashboard can trigger a cascade of network requests across multiple services, databases, and third-party APIs [1]. Security operations centers increasingly rely on artificial intelligence to manage the scale and complexity of modern enterprise environments. AI-assisted dashboards aggregate telemetry from heterogeneous sources and apply machine learning models to detect anomalies, assess risk, and recommend actions. Despite improvements in model performance, human analysts remain responsible for interpreting outputs and making final decisions. This human–AI partnership introduces a critical challenge: ensuring that analysts appropriately calibrate their trust in AI-generated insights.

When a failure occurs, the symptom is often decoupled from the root cause. A dashboard might fail to load a widget, displaying a generic "500 Internal Server Error," while the actual issue lies in a schema mismatch between two internal services or a silent browser-side JavaScript exception.

Current industry standard tools—such as Application Performance Monitoring (APM) agents and centralized logging systems—excel at capturing raw telemetry. However, they typically present this data in silos. A support engineer must manually toggle between browser console logs, API gateways, and backend server logs to reconstruct the timeline of a failure [2]. This manual correlation is time-consuming, error-prone, and requires deep domain knowledge.

Furthermore, the output of these tools is highly technical, meant for developers rather than end-users or Level 1 support staff. There is a critical gap in the current ecosystem for a system that can:

1. **Unified Data Collection:** Aggregating data from the client, network, and server layers simultaneously.
2. **Automated Correlation:** Linking these disparate data points to a single user session or transaction ID.
3. **Semantic Explanation:** Translating the correlated technical data into natural language that explains *what* went wrong and *why*.

To address these challenges, we present a framework for an **Automated Multi-Source Debugging Engine**. This system leverages recent advancements in Generative AI to synthesize technical logs into human-readable explanations, streamlining the debugging process and democratizing system observability.

## II. Related Work

A. Distributed Tracing and Observability

Distributed tracing tools like Jaeger and Zipkin have become standard for visualizing requests across microservices. While effective for latency analysis and identifying failing nodes, they often lack the granularity of browser-side events and do not validate the content of the data against defined schemas [3]. Our approach builds upon tracing by integrating client-side context and schema validation.

B. Automated Log Analysis

Automated log analysis using machine learning (AIOps) has been widely studied for anomaly detection. Techniques involving clustering (e.g., LogCluster) group similar error patterns to reduce noise [4]. However, these systems largely remain "human-in-the-loop" for interpretation. They identify that an anomaly occurred but rarely explain why in natural language.

C. Large Language Models in Software Engineering

The application of LLMs to software engineering tasks, such as code generation and bug fixing, is a rapidly growing field. Recent studies have demonstrated the capability of models like GPT-4 to explain code snippets and suggest fixes [5]. This paper extends this capability to the domain of runtime error explanation, using the model not just to analyze static code, but to interpret dynamic runtime states.

### III. System Architecture

The Adaptive Trust Visualization framework operates as a conceptual layer between backend analytics and front-end rendering. Its primary function is to translate trust assessments into perceptual cues within the user interface. The proposed system architecture [Fig. 1] is designed to sit as an observability layer between the client application and the backend infrastructure. It consists of four primary modules: (1) Multi-Source Data Collector, (2) API Contract Validator, (3) Event Correlation Engine, and (4) Natural Language Generation Module.

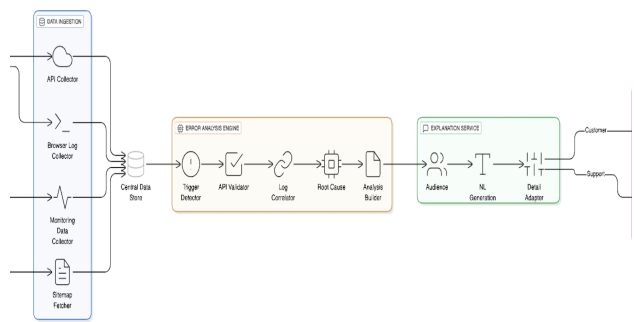

Fig. 1 System Architecture

**A. Trust Signal Modeling**

Trust is modeled as a multi-dimensional construct derived from heterogeneous signals. Quantitative inputs may include model confidence estimates, data freshness, source reliability, and correlation strength across signals.

Historical validation outcomes, such as prior false positives or confirmed detections, also contribute to trust estimation.

Each signal is normalized to a common scale and weighted to reflect its relative importance. The resulting composite trust representation provides a continuous measure of reliability rather than a binary classification. In addition to numerical aggregation, semantic reasoning is applied to contextual information such as system summaries or analyst feedback. This qualitative interpretation allows the framework to adjust trust assessments based on factors that are difficult to capture numerically.

By combining quantitative and qualitative signals, the framework achieves robust trust modeling while remaining agnostic to specific data sources or proprietary implementations.

**B. Adaptive Rendering Techniques**

The adaptive rendering layer translates trust assessments into perceptual cues embedded in the user interface. Rather than introducing explicit labels or alerts, trust is conveyed through subtle visual modulation. Common techniques include adjustments to opacity, saturation, brightness, and visual clarity.

High-trust elements are rendered with greater visual prominence, appearing crisp and fully saturated. Lower-trust elements are subtly de-emphasized through translucency or desaturation, signaling uncertainty without removing access to information. These cues are designed to be pre-attentive, allowing analysts to rapidly prioritize information without conscious effort.

Visualization literature suggests that pre-attentive cues can significantly reduce cognitive load and improve task performance [6], [7]. By embedding trust directly into visual semantics, the framework enhances usability while avoiding additional interface complexity.

**C. Multi-Source Data Collector**

The foundation of the system is the ability to capture the state of the application at the precise moment of failure. The collector operates on three distinct planes:

- **Browser Plane:** A lightweight JavaScript SDK intercepts console errors, unhandled promise rejections, and network stack traces. It captures the user's navigational context (e.g., current URL, clicked element) to provide reproducibility.
- **Network Plane:** The system inspects HTTP request and response headers, payloads, and status codes. Crucially, it captures the "har" (HTTP Archive) data surrounding a failure.
- **Server Plane:** Through middleware integration, the collector ingests stack traces and application logs generated during the processing of the specific request ID associated with the client action.

## D. API Contract Validator

A common source of dashboard failures is "drift" between the API consumer (frontend) and the API provider (backend). The Contract Validator module loads the formal API specification (e.g., OpenAPI/Swagger definitions) and performs real-time validation of traffic.

When an error occurs, the system checks if the request payload sent by the browser matched the expected schema, and if the response from the server adhered to the defined structure. This allows the system to instantly classify errors as "Client-Side Schema Violation" or "Server-Side Contract Breach," saving hours of debugging time.

## E. Event Correlation Engine

Raw data from the collector is often noisy. The Correlation Engine utilizes a time-windowed algorithm and Trace ID injection to align events.

Let $E_b$, $E_n$, and $E_s$ represent sets of events from the Browser, Network, and Server, respectively. The engine seeks to find a subset of events $S \subset (E_b \cup E_n \cup E_s)$ such that all events in $S$ share a unique correlation identifier $C_{id}$ or fall within a correlation window $\Delta t$ centered around the user-reported failure time $T_f$.

$$S \subset \{e \mid \texttt{timestamp (e)} \in [T_f - \Delta t, T_f + \Delta t] \wedge \texttt{match (e, } C_{id})\}$$

This filtering process isolates the signal from the noise, creating a unified "Failure Context Object" (FCO).

## F. Natural Language Generation (NLG) Module

The FCO, containing the technical root cause (e.g., a stack trace or validation error), is fed into the NLG module. This module utilizes a fine-tuned Large Language Model (LLM). The model is prompted with the FCO and instructed to:

1. Summarize the technical error.
2. Identify the likely culprit (Database, Network, Client logic).
3. Generate a user-facing explanation (e.g., "The dashboard failed to load because the reporting service is unavailable").
4. Generate a developer-facing explanation (e.g., "GET /reports failed with 500 Error due to SQL timeout in the analytics microservice").

## IV. Methodology and Implementation

To validate this architecture, we define a processing pipeline that activates upon exception detection.

1. Trigger Mechanism:

The process is triggered either automatically (upon detection of a 4xx/5xx status code) or manually (user clicks "Report Issue").

2. Context Assembly:

The system snapshots the Redux/State store (sanitized for PII) and retrieves the relevant logs. A key innovation here is the Sanitization Layer, which uses Named Entity Recognition (NER) to mask sensitive data (IP addresses, user IDs) before the data is passed to the analysis engine or LLM, ensuring enterprise-grade security.

3. Heuristic Analysis:

Before invoking the LLM, a heuristic analyzer checks for known patterns.

- *Scenario A:* If the API Contract Validator reports a "Required Field Missing" error, the system deterministically flags this as a Frontend Bug.
- *Scenario B:* If the Network Plane reports a timeout, it is flagged as an Infrastructure Issue.

4. Generative Explanation:

If heuristic analysis is insufficient, the aggregated context is serialized into a prompt for the LLM.

- *Input:* JSON object containing {Endpoint: "/api/v1/data", Status: 500, Log: "NullPointerException at Controller.java:45", Browser: "React Error Boundary caught error"}.
- *Output Generation:* The model synthesizes this into a narrative: "The application crashed because the backend received unexpected null data for the requested chart, causing a server-side exception."

## V. Discussion and Use Cases

A. Improved Support Efficiency

In a traditional workflow, a support engineer receiving a ticket stating "The dashboard is broken" must initiate a back-and-forth conversation to get screenshots and logs. With the proposed system, the ticket is auto-populated with the generated explanation and the correlated technical FCO. This eliminates the "discovery" phase of support, allowing engineers to move immediately to remediation.

B. End-User Experience

Replacing "An unknown error occurred" with "We are having trouble connecting to the inventory database, please try again in 5 minutes" significantly reduces user frustration. It builds trust by being transparent about the nature of the failure.

C. Cross-Team Collaboration

The API Contract Validation feature acts as an arbiter between Frontend and Backend teams. By automatically identifying which side violated the contract, the system reduces the "blame game" often seen during incident response.

## VI. Theoretical Validation

While a full-scale production deployment analysis is outside the scope of this paper, we estimate the impact based on standard incident response metrics.

The breakdown of Mean Time to Resolution (MTTR) generally consists of:

$$MTTR = T_{detect} + T_{diagnose} + T_{fix}$$

Our system targets $T_{diagnose}$. In complex microservices, diagnosis often consumes 60-70% of the total resolution time [8]. By automating the correlation of logs and contract validation, we hypothesize a reduction in $T_{diagnose}$ by approximately 40-50%, as the manual step of log hunting is removed.

## VII. Conclusion

The complexity of modern dashboard applications demands a new generation of debugging tools. We have proposed an automated system that bridges the gap between raw telemetry and human understanding. By combining multi-source data collection, strict API contract validation, and Generative AI, we can transform opaque error signals into clear, actionable intelligence. This approach not only streamlines engineering operations but also enhances the overall reliability and user experience of enterprise software. Future work will focus on integrating "Self-Healing" capabilities, where the system not only explains the error but automatically attempts remediation strategies based on the diagnosis.

## References

[1] M. Fowler and J. Lewis, "Microservices," IEEE Software, vol. 31, no. 6, 2014.

[2] R. Chow, "The State of Observability 2023," ACM Queue, vol. 21, no. 2, pp. 15-28, 2023.

[3] B. Sigelman et al., "Distributed Tracing in Practice," O'Reilly Media, 2020.

[4] S. He et al., "Experience Report: System Log Analysis for Anomaly Detection," 2016 IEEE 27th International Symposium on Software Reliability Engineering (ISSRE), Ottawa, ON, Canada, 2016.

[5] J. Zhang, "A Survey of Large Language Models for Code," arXiv preprint arXiv:2306.02345, 2023.

[6] T. Munzner, Visualization Analysis and Design. Boca Raton, FL, USA: CRC Press, 2014.

[7] A. Endert et al., "Human–AI Interaction in Visual Analytics," IEEE Computer Graphics and Applications, vol. 39, no. 3, pp. 27–41, 2019.

[8] "The Cost of Downtime," Gartner IT Operations Management Report, 2022.